\documentclass[twocolumn,aps,prc,floatfix]
{revtex4}

\usepackage[utf8]{inputenc}
\usepackage{bbm}
\usepackage{amssymb}
\usepackage{amsmath}
\usepackage{graphicx}
\usepackage[normalem]{ulem}
\usepackage[dvips]{color}
\usepackage{appendix}
\usepackage{color}

\def\es0{$E_{\rm sym}(\rho_0)$}

\def\us0{$U_{\rm sym}(\rho_0,k_F)$~}

\def\Ka{$K_{\rm A}$~}

\def\Kt{$K_{\tau}$~}
\def\Ki{$K_{\infty}$~}

\begin{document}
\title{Differential analysis of incompressibility in neutron-rich nuclei}

\author{Bao-An Li$^{1}$\footnote{Corresponding author: Bao-An.Li@Tamuc.edu} and Wen-Jie Xie$^{2}$}
\address {$^{1}$Department of Physics and Astronomy, Texas A\&M University-Commerce, TX 75429-3011, USA}
\address {$^{2}$Department of Physics, Yuncheng University, Yuncheng 044000, China}

\date{\today}

\begin{abstract}
Both the incompressibility \Ka of a finite nucleus of mass A and that ($K_{\infty}$) of infinite nuclear matter are fundamentally important for many critical issues in nuclear physics and astrophysics. While some consensus has been reached about the $K_{\infty}$, accurate theoretical predictions and experimental extractions of $K_{\tau}$ characterizing the isospin dependence of \Ka have been very difficult. We propose a differential approach to extract the \Kt and \Ki independently from the \Ka data of any two nuclei in a given isotope chain. Applying this new method to the \Ka data from isoscalar giant monopole resonances (ISGMR) in even-even Pb, Sn, Cd and Ca isotopes taken by U. Garg {\it et al.} at the Research Center for Nuclear Physics (RCNP), Osaka University, Japan, we find that  the $^{106}$Cd-$^{116}$Cd and $^{112}$Sn-$^{124}$Sn pairs having the largest differences in isospin asymmetries in their respective isotope chains measured so far provide consistently the most accurate up-to-date \Kt value of $K_{\tau}=-616\pm 59$ MeV and $K_{\tau}=-623\pm 86$ MeV, respectively, largely independent of the remaining uncertainties of the surface and Coulomb terms in expanding the $K_{\rm A}$, while the $K_{\infty}$ values extracted from different isotopes chains are all well within the current uncertainty range of the community consensus for $K_{\infty}$.
Moreover, the size and origin of the ``Soft Sn Puzzle" is studied with respect to the ``Stiff Pb Phenomenon". It is found that the latter is favored due to a much larger (by $\sim 380$ MeV) \Kt for Pb isotopes than for Sn isotopes,
while the \Ki from analyzing the \Ka data of Sn isotopes is only about 5 MeV less than that from analyzing the Pb data.
\end{abstract}

\maketitle

\section{Introduction.}
Because of its fundamental importance in nuclear physics and broad impacts on astrophysics, the incompressibility $K_{\infty}$ of infinite nuclear matter has been a long standing and major scientific goal of many experimental and theoretical researches. Since the pioneering work of Blaizot who determined $K_{\infty}$ = (210 $\pm$ 30) MeV from analyzing the experimental data on giant monopole resonance (GMR) energies in $^{\rm 40}$Ca, $^{\rm 90}$Zr and $^{\rm 208}$Pb \cite{Blaizot}, extensive theoretical studies and systematic experiments on the incompressibility \Ka of finite nuclei extracted from GMR energies over the last four decades \cite{Blaizot,You,Garg18,Jorge10,Stone,Colo14} have led to the community consensus that the $K_{\infty}$ is in the range of 220 MeV to 260 MeV \cite{Garg18,Colo14,shlomo06} or around
$235\pm 30$ MeV \cite{Khan1,MM1}. Thanks to the new advancement in experiments especially at rare isotope beam facilities, GMR energies of neutron-rich nuclei along long isotope chains have become possible recently, facilitating more accurate and extensive explorations of the isospin dependence of $K_{\rm A}$.

The incompressibility $K_{\rm A}$ of finite nuclei is usually parameterized in the form of a leptodermous expansion in powers of A$^{\rm -1/3}$ in typical macroscopic models as \cite{Blaizot}
\begin{equation}\label{KA-e}
K_A \approx K_\infty(1+cA^{-1/3})+K_\tau \delta^2+K_{\rm {Cou}}Z^2A^{-4/3}
\end{equation}
for a nucleus of mass number A, charge number Z and isospin asymmetry $\delta=\frac{N-Z}{A}$, with $c\approx -1.2\pm 0.12$ \cite{Pat02} and $K_{\rm Cou}\approx -5.2 \pm 0.7$  MeV \cite{Sag07} being the surface and Coulomb parameters, respectively. The $K_{\tau}$ characterizing the isospin dependence of \Ka has been the main focus of many recent experimental and theoretical investigations. By moving the Coulomb term to the left side of the above equation, {\it for all practical purposes} \cite{Garg18} in extracting the
\Kt from the experimental \Ka data \cite{Li07,Li10,Pat12,Pat13,How19,How20}, the $K_A -K_{\rm{Cou}}Z^2A^{-4/3}$ was fitted with a quadratic function of the form ${\bf a}+ K_\tau \delta^2$ assuming ${\bf a} = K_{\infty}(1 + cA^{-1/3})$ is a constant. This approach resulted in an ``experimental'' value of $K_{\tau}$ = -550$\pm$100 MeV from the \Ka data of Sn isotopes and $K_{\tau}$ = -555$\pm$75 MeV from the Ca isotopes, respectively. The mass dependence of ${\bf a}$ and the known correlation between \Ki and \Kt neglected in the above approach were found to affect significantly the extracted \Kt values \cite{Stone,Pea10}. For example, using the same $c$ and $K_{\rm Cou}$ parameters but preserving the mass dependence of ${\bf a}$ and considering the correlation between \Ki and \Kt in the error minimization of a multivariate $\chi^2$ fit, $K_{\tau}$ = -595$\pm$177 MeV, \Ki=$209\pm 6$ MeV from Sn isotopes, and $K_{\tau}$ = -463$\pm$405 MeV,  \Ki=$211\pm 11$ MeV from the Cd isotopes were found \cite{Stone}. As it was stressed already \cite{Garg18,Stone,Pea10,Shlomo}, the state of affairs in understanding and extracting the \Kt has been very unsatisfactory for a long time. 

While it is well known that the \Ki is a fundamental quantify critical for solving many interesting issues in both nuclear physics and astrophysics, to our best knowledge, 
impacts of \Kt on astrophysical observables, e.g., radii of neutron stars, are only indirect through the shared underlying isovector interactions. Nevertheless, an accurate value of \Kt is useful for predicting the incompressibilities and thus the collective excitations of heavy neutron-rich nuclei that have not been measured or can not be measured directly because of their instabilities. It is thus imperative to find more robust methods to extract accurately both the \Ki and especially the \Kt from the \Ka data. Such methods are also expected to play important roles in analyzing the coming new data from measuring the \Ka of exotic, more neutron-rich nuclei in long isotopic chains at advanced radioactive beam facilities.

In some earlier studies, see, e.g., Ref. \cite{Garg18} for a recent review, the leptodermous expansion of Eq. (\ref{KA-e}) was used to extract its coefficients by performing $\chi^2$ fittings to the experimental \Ka data. We regard this approach as the integral approach in our following discussions. It has been shown in numerous works, see, e.g., \cite{Vinas}, that such approach is not very accurate.  It was concluded that these leptodermus coefficients  are not well constrained by the experimental data. Later, some sort of community ``consensus" was reached that the model to analyze \Ki must contain microscopic effects, reproduce the GMR and also other observables before one extrapolates to the infinite system. In reality, in analyzing the GMR data from RCNP, for instance, the ``consensus" approach was used in extracting only the \Ki and indeed much interesting physics was obtained. However, the ``experimental" \Kt value was always extracted from the same \Ka data using the Eq. (\ref{KA-e})  \cite{Li07,Li10,Pat12,Pat13,How19,How20} because the \Kt is only defined through this equation for the incompressibility of finite nuclei.
Thus, regardless of whatever criticisms people may have for using the Eq. (\ref{KA-e}) to extract the $K_{\infty}$, the same ``consensus" approach does not apply to the extraction of \Kt which can only be extracted by using the  Eq. (\ref{KA-e}). Moreover, given the still very large dependences on both the many-body theories and interactions used in the  ``consensus" approach in studying the \Ki for infinite nuclear matter, some of the same techniques can not be used in calculating the \Kt for finite nuclei and the relevant isovector interactions are much less known than the isoscalar interactions. 

Indeed, shell and pairing effects are not considered in the Eq. (\ref{KA-e}). These effects may play some roles in extracting the incompressibility from ISGMR data \cite{Ehan}, but their effects are still much smaller than the current uncertainty range of the fiducial value of \Ki not to mention the huge uncertainty of \Kt discussed above when these effects are neglected. We notice that it was already pointed out that shell effects are not important for ISGMR \cite{Pat13} as giant resonances are basically collective phenomena. Obviously, more researches are necessary to quantify more precisely the shell and pairing effects on ISGMR. 

The well-known problems mentioned above about the reliability of the leptodermus coefficients extracted from the $\chi^2$ fitting do not necessarily mean that the Eq. (\ref{KA-e}) itself is wrong or inaccurate, they may indicate instead that the $\chi^2$ fitting approach is not appropriate for extracting the \Ki and \Kt values. Thus, they do not prevent people from using the same Eq. (\ref{KA-e}) in better or more appropriate ways to extract accurately the \Ki and \Kt from the same \Ka data.

In this work, we propose a differential approach to extract exactly the values of \Ki and \Kt independently from the \Ka data of two nuclei in any isotopic chain. The nucleus-nucleus pair having the largest difference in their isospin asymmetries is found to give the most accurate \Kt and \Ki values simultaneously. Effects of varying the $c$ and $K_{\rm Cou}$ parameters by $\pm20\%$ around their known most probable values on extracting both \Kt and \Ki are also examined. While the variations of $c$ and $K_{\rm Cou}$ lead the extracted \Ki values to vary within its current consensus range, they have almost no effect on extracting the $K_{\tau}$, indicating the robustness of the differential approach. We found that both the mean value and uncertainty we extracted for \Ki are compatible with those from using the state-of-the-art microscopic theories in the ``consensus" approach, while the accuracy of the extracted \Kt in our approach is much higher than what is available in the literature.  Finally, it has been known for about a decade that the \Ka values extracted experimentally from Sn isotopes are apparently smaller compared to predictions of non-relativistic mean-field or relativistic mean-field + Random Phase Approximation (RPA) calculations that can successfully describe the ISGMR data of Pb isotopes \cite{Li10}.  However, the origin of this so-called ``Soft Sn Puzzle" \cite{Garg18,Jorge09} or ``Stiff Pb Phenomenon" \cite{Ehan} is still unclear. We shall investigate if the differential analysis can shed new light on this issue.

The rest of the paper is organized as follows. In the next section, we present details of the proposed differential analyses. In section \ref{analysis}, we perform a differential analysis for the \Ka data  \cite{Li07,Li10,Pat12,Pat13,How19,How20} from isoscalar giant monopole resonances in even-even Pb, Sn, Cd and Ca isotopes taken by U. Garg {\it et al.} at RCNP.  In section \ref{surf}, we study effects of the remaining uncertainties of the surface and Coulomb parameters on extracting the \Ki and \Kt values. Finally, we summarize and draw conclusions of our work.

\section{The differential approach} Applying the Eq.(\ref{KA-e}) to any two isospin asymmetric ($\delta\neq 0$) nuclei of mass and charge $(A_1,Z_1)$ and $(A_2,Z_2)$ separately, the \Kt and \Ki can be expressed exactly as
\begin{widetext}
\begin{eqnarray}\label{KTI}
&&K_{\tau} =\left[\frac{K_{\rm A_1}}{S_1}-\frac{K_{\rm A_2}}{S_2}-K_{\rm{Cou}}\left(\frac{Z_1^2A_1^{-4/3}}{S_1}-\frac{Z_2^2A_2^{-4/3}}{S_2}\right)\right]
\bigg/\left(\frac{\delta_1^2}{S_1}-\frac{\delta_2^2}{S_2}\right),\\
&&K_{\infty} =\left[\frac{K_{\rm A_1}}{\delta_1^2}-\frac{K_{\rm A_2}}{\delta_2^2}-K_{\rm{Cou}}\left(\frac{Z_1^2A_1^{-4/3}}{\delta_1^2}-\frac{Z_2^2A_2^{-4/3}}{\delta_2^2}\right)\right]
\bigg/\left(\frac{S_1}{\delta_1^2}-\frac{S_2}{\delta_2^2}\right)
\end{eqnarray}
\end{widetext}
where $S_i=1+cA_i^{-1/3}$ for the nucleus-i with i=1 or 2.  One can understand intuitively the physical meanings of the above expressions by using the mathematical definitions of \Kt and $K_{\infty}$ based on Eq. (1). Namely, neglecting the Coulomb correction, $K_{\tau}\equiv\left(\partial K_{\rm A}/\partial \delta^2\right)_{\rm S}\approx \Delta (K_{\rm A}/S)/\Delta(\delta^2/S)=(\frac{K_{\rm A_1}}{S_1}-\frac{K_{\rm A_2}}{S_2})/(\frac{\delta_1^2}{S_1}-\frac{\delta_2^2}{S_2})$ gives the leading term of $K_{\tau}$ in Eq. (2). It is simply the changing rate of \Ka with respect to $\delta^2$ evaluated by using the ratio of their finite changes.
Similarly, $K_{\infty}\equiv\left(\partial K_{\rm A}/\partial S\right)_{\delta}\approx \Delta (K_{\rm A}/\delta^2)/\Delta(S/\delta^2)=(\frac{K_{\rm A_1}}{\delta_1^2}-\frac{K_{\rm A_2}}{\delta_2^2})/(\frac{S_1}{\delta_1^2}-\frac{S_2}{\delta_2^2})$ gives the leading term of $K_{\infty}$ in Eq. (3).

We notice that while the \Kt and \Ki are determined independently by the \Ka data themselves of any two nuclei used, they satisfy the constraint given by Eq. (\ref{KA-e}). Therefore, there is an intrinsic correlation between \Kt and \Ki when they are varied using the \Ka data of many different nucleus-nucleus pairs in a given isotopic chain as we shall demonstrate.

The corresponding uncertainties of $K_{\tau}$ and $K_{\infty}$ can be calculated
exactly according to the rules of error propagation using the experimental errors of \Ka data, i.e., $\sigma_{K_{\rm A_1}}$ and $\sigma_{K_{\rm A_2}}$ in the nucleus-1 and nucleus-2 considered.
Nevertheless, to see analytically what nucleus-nucleus pairs may give the most accurate \Kt and \Ki values, we notice that for heavy nuclei in the same isotope chain, $S_1\approx S_2\approx 1$, the error bars are reduced to
\begin{eqnarray}\label{KT-e}
\sigma_{K_{\tau}} &\approx&\sqrt{\sigma_{K_{\rm A_1}}^2+\sigma_{K_{\rm A_2}}^2}\bigg/\bigg|\delta_1^2-\delta_2^2\bigg|,\\
\sigma_{K_{\infty}} &\approx&\sqrt{(\delta_2^2\cdot\sigma_{K_{\rm A_1}})^2+(\delta_1^2\cdot\sigma_{K_{\rm A_2}})^2}\bigg/\bigg|\delta_1^2-\delta_2^2\bigg|.
\end{eqnarray}
They both are {\it inversely} proportional to $|\delta_1^2-\delta_2^2|$, thus nuclear pairs having the largest difference in their isospin asymmetries will give the most accurate \Kt and \Ki values simultaneously. Moreover,
because of the weighting of $\sigma_{K_{A}}$ by $\delta^2\ll1$ in evaluating the $\sigma_{K_{\infty}}$, the \Ki can be more precisely evaluated than the $K_{\tau}$, explaining the relatively larger errors of the extracted \Kt values.

While in principle the above formalisms can be applied to any two nuclei, we shall restrict their applications to nuclei in the same isotopic chain. This will reduce not only effects of systematic experimental errors as what is being
used is the {\it difference} in \Ka scaled by either the surface factor S or isospin asymmetry $\delta$ of the two nuclei in the same isotopic chain, but also effects of the higher-order terms neglected in expanding the \Ka in Eq. (\ref{KA-e}). This is also one of the reasons why the differential approach can more precisely extract both the \Kt and \Ki compared to typical integral approaches normally used in the literature. 

In cases where one of the nuclei is isospin-symmetric, say $\delta_1=0$, its \Ka alone can be used to evaluate the \Ki according to $K_{\infty} =K_{\rm A_1}/S_1-K_{\rm{Cou}}Z_1^2A_1^{-4/3}/S_1$ while the \Kt can be evaluated from the Eq. (2) by choosing the nucleus-2 as neutron-rich as possible to get the most accurate result, indicating the importance of using exotic heavy isotopes. As noticed already in the literature, see, e.g., Ref. \cite{Shlomo}, the leptodermous expansion in Eq. (1) itself may not be a good approximation for light nuclei, the differential approach is thus expected to work better for more heavy nuclei.

\section{Differential analyses of the \Ka data from experiments at RCNP}\label{analysis}
\begin{figure*}[htb]
\begin{center}
\resizebox{0.8\textwidth}{!}{
 \includegraphics[width=\linewidth]{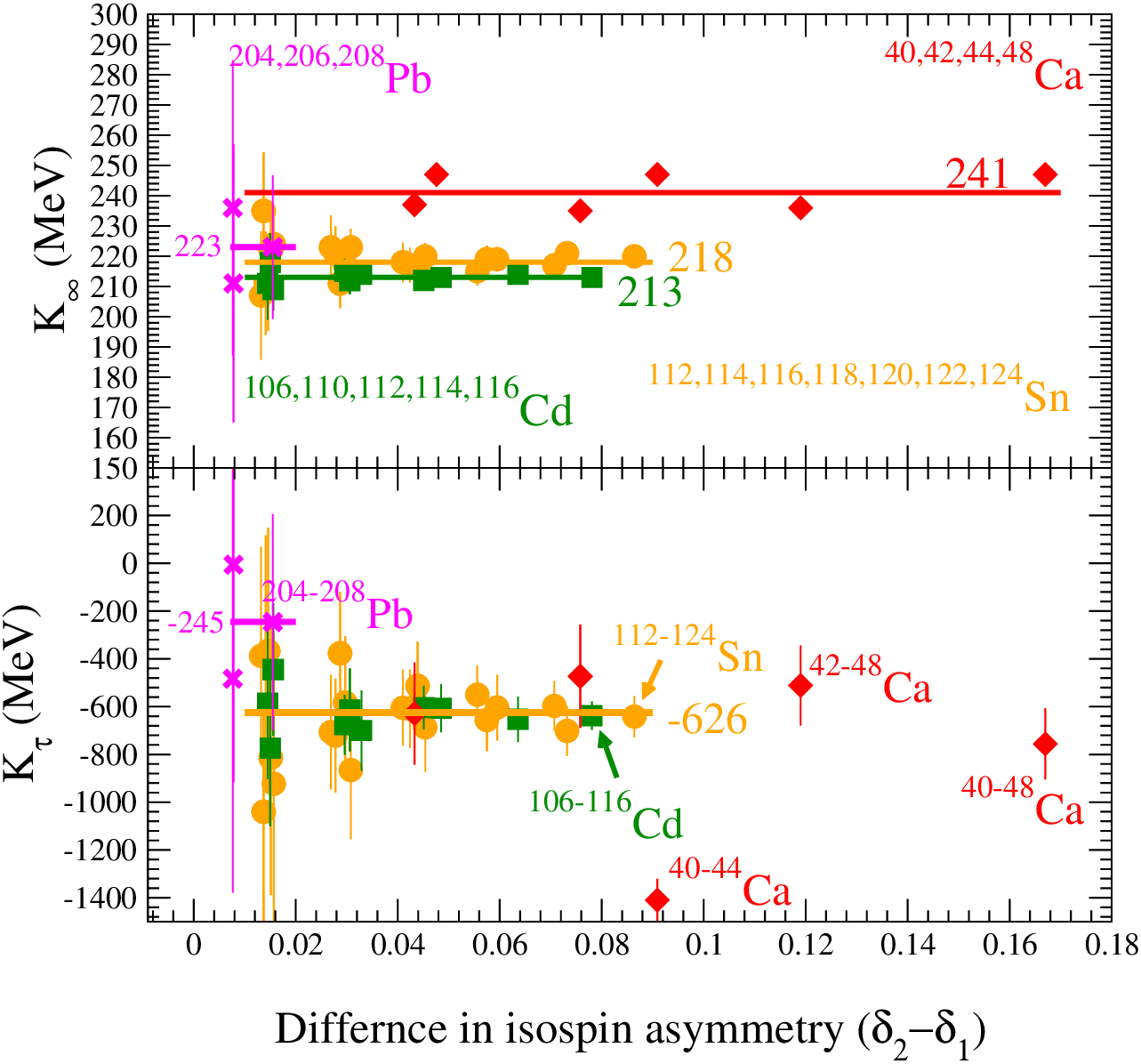}
 }
  \caption{(Color online) The \Kt (lower window) and \Ki (upper window) from differential analyses of the incompressibilities in finite nuclei as functions of the difference ($\delta_2-\delta_1$) in isospin asymmetries of the isotope pairs used. The solid lines are the mean values of \Kt and \Ki for the respective isotope chains. The arrows indicate the Cd and Sn isotope pairs giving the most accurate \Kt and \Ki values.}\label{figure1}
\end{center}
\end{figure*}

Shown in Fig. \ref{figure1} are the results of our differential analyses of the \Ka data in $^{204,206,208}$Pb,
$^{112,114,116,118,120,122,124}$Sn, $^{106,110,112,114,116}$Cd and $^{40,42,44,48}$Ca from the GMR experiments at RCNP \cite{Li07,Li10,Pat12,Pat13,How19,How20} using $c=-1.2$ and $K_{\rm Cou}=-5.2$ MeV. The extracted \Kt and \Ki values are shown as functions of the difference ($\delta_2-\delta_1$)
in isospin asymmetries of the two nuclei involved in each isotope chain.
Except for the Pb isotopes, we took the \Ka data directly from the experimental publications as listed in Table \ref{T1}. They derived the \Ka values using the moment ratios for the ISGMR energies $E_{\rm{ISGMR}}$ and the experimental charge radii $\sqrt{<r^2>}$ from Ref. \cite{Cradius} according to the relation
\begin{equation}
K_{\rm{A}}=\left(\frac{E_{\rm{ISGMR}}}{\hbar c}\right)^2 Mc^2 <r^2>
\end{equation}
where M is the average nucleon mass. They did not publish the \Ka values for the three Pb isotopes. We derived their \Ka values using the published
ISGMR energies (from $\sqrt{m_1/m_{-1}}$) \cite{Pat13} and their charge radii from Ref. \cite{Cradius}. More quantitatively, we found that $K_{\rm A}$ is $136.93 \pm 1.99$ MeV, $137.44 \pm 1.99$ MeV and $136.44 \pm 1.99$ MeV, respectively, for $^{204}$Pb, $^{206}$Pb and $^{208}$Pb.
\begin{table}\small
  \caption{The incompressibility data of finite nuclei analyzed.}
  \vspace{0.6cm}
    \begin{tabular}{|c|c|c|}
   \hline
   Nucleus & $K_A$ (MeV)  & Reference \\
       \hline
    $^{40}$Ca & $144.46 \pm 0.33$ & \cite{How20}\\
    $^{42}$Ca & $139.00 \pm 1.09$ & \\
    $^{44}$Ca& $137.36 \pm 0.66$ & \\
    $^{48}$Ca& $131.90 \pm 4.13$ & \\
        \hline
       \hline
      $^{106}$Cd & $127.84 \pm 0.86$ & \cite{Pat12}\\
    $^{110}$Cd & $124.59 \pm 0.86$ & \\
    $^{112}$Cd & $123.59 \pm 0.77$ & \\
    $^{114}$Cd & $120.95 \pm 1.24$ & \\
    $^{116}$Cd & $118.96 \pm 0.86$ & \\
     \hline
   \hline
    $^{112}$Sn & $131.86 \pm 1.53$ & \cite{Li07,Li10}\\
    $^{114}$Sn & $129.45 \pm 1.64$ & \\
    $^{116}$Sn & $127.11 \pm 1.53$ & \\
    $^{118}$Sn & $126.39 \pm 1.54$ & \\
    $^{120}$Sn & $125.45 \pm 1.63$ & \\
    $^{122}$Sn & $121.33 \pm 1.54$ & \\
    $^{124}$Sn & $120.17 \pm 1.62$ & \\
   \hline
   \hline
    $^{204}$Pb & $136.93 \pm 1.99$ &  \cite{Pat13}\\
    $^{206}$Pb & $137.44 \pm 1.99$ &  \\
    $^{208}$Pb & $136.44\pm 1.99$ &  \\
     \hline
   \hline
       \end{tabular}
  \label{T1}
\end{table}

Several interesting observations can be made: (1) The uncertainties of both \Kt and \Ki generally decreases while their mean values remain approximately constants with the increasing $(\delta_2-\delta_1)$ for each isotope chain. (2) The $^{106}$Cd-$^{116}$Cd and $^{112}$Sn-$^{124}$Sn pairs give the most accurate and consistent values of $K_{\tau}=-616\pm 59$ MeV and $K_{\tau}=-623\pm 86$ MeV, respectively. (3) The \Kt values from analyzing the relatively light $^{40,42,44,48}$Ca isotopes have larger error bars and scatter around broadly at small isospin separations. However, they seem to converge at large isospin separations and become generally consistent with the means from analyzing the Sn and Cd isotopes within error bars. We notice that among all data available from the RCNP experiments, the $^{40-48}$Ca pair has the highest isospin separation $(\delta_2-\delta_1)=0.167$. This pair gives $K_{\tau}=-756\pm 149$ MeV.  As mentioned earlier, the \Ka expansion of Eq. (1) is not expected to work well for light nuclei. The scattering of the \Kt values from analyzing the Ca data may thus indicate that our differential approach based on Eq. (1) has reached its validity limit. (4) The $^{106}$Cd-$^{116}$Cd and $^{112}$Sn-$^{124}$Sn pairs also give the most accurate \Ki values of  \Ki=$213\pm 2$ MeV and \Ki=$220\pm 3$ MeV, respectively. (5) The extracted \Ki shows the well-known isotope dependence found earlier when the Eq. (1) was used previously in $\chi^2$ fittings of the \Ka data \cite{Blaizot,Garg18,Jorge10,Stone,Colo14,Shlomo} although our differential approach does not use any fitting at all. Nevertheless, the variation of the \Ki from Cd to Ca isotopes is well within the uncertainty range of the currently reached consensus value for $K_{\infty}$.

While the three Pb isotopes pairs have very small isospin separations of 0.00765, 0.00781 and 0.0155, respectively, the results from the differential analyses of their \Ka values set a useful reference for comparisons and favor a ``Stiff Pb Phenomenon" \cite{Ehan} instead of the so-called ``Soft Sn Puzzle" existing in the literature. The Pb data give an average value of $K_{\infty}=223.1 \pm 39.5$ MeV and $K_{\tau}=-245 \pm 753$ MeV, respectively. Their means are indicated by the horizontal magenta bars for comparisons. Two important indications are worth emphasizing. Firstly, it was not known before what is the cause of the ``Soft Sn Puzzle". Our analysis here indicates that the mean value of $K_{\infty}$ from Sn isotopes (218 MeV) is only 5 MeV below that from Pb isotopes (223 MeV) well within the experimental uncertainties. This finding happens to be the same as that found in a very recent Bayesian uncertainty quantification of the nuclear matter incompressibility using the original GMR data of the same sets of isotopes analyzed within the Skyrme Hartree-Fock plus RPA approach \cite{Xu21}. On the other hand, the mean value of $K_{\tau}$ from Sn isotopes (-626 MeV) is significantly below that (-245 MeV) from the Pb isotopes although the latter also has a large error bar. It indicates that the ``Soft Sn Puzzle" is mainly due to the significantly smaller \Kt value for Sn isotopes or larger \Kt value for Pb isotopes. The overpredictions of the \Ka values by the state-of-the-art microscopic theories are most likely due to the model ingredients controlling the \Kt instead of the \Ki values. Secondly, 
it is interesting to note that the \Kt values from Sn, Cd and Ca isotopes all converged asymptotically at large isospin separations to relatively precise values with less than about 20\% errors. However, the average \Kt value for Pb isotopes is significantly higher than these asymptotic values although it is only slightly higher than the \Kt values for Sn and Cd isotopes at the same small isospin separations. These findings may give us some hints about whether 
there is a ``Soft Sn Puzzle" or a ``Stiff Pb Phenomenon". Our results seem to indicate that it is probably more meaningful to speak about a `Stiff Pb Phenomenon". To verify the latter experimentally, more \Ka data for Pb isotopes are obviously necessary. 
 
 \begin{figure}[htb]
\vspace{0.55cm}
\begin{center}
 \resizebox{0.48\textwidth}{!}{
  \includegraphics{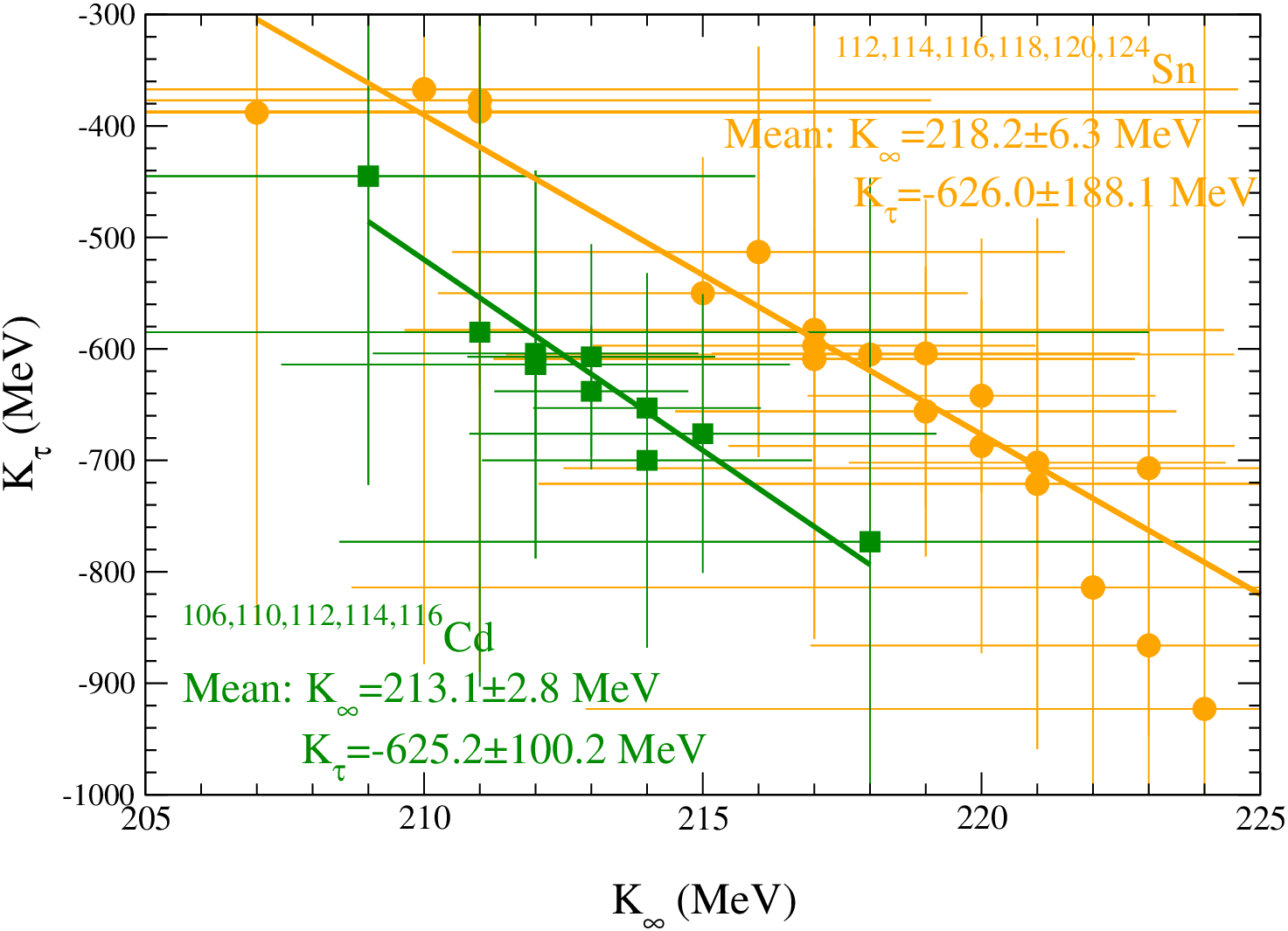}}
  \caption{(Color online) The correlation between \Kt and \Ki with each point representing one nucleus-nucleus pair in the Cd or Sn isotope chain corresponding to the results shown in Fig. \ref{figure1}. The solid lines are results of a $\chi^2$ fit to all points in each isotope chain. }\label{figure2}
\end{center}
\end{figure}
\begin{figure*}[htb]
\begin{center}
 \resizebox{1.0\textwidth}{!}{
 \includegraphics[width=\linewidth]{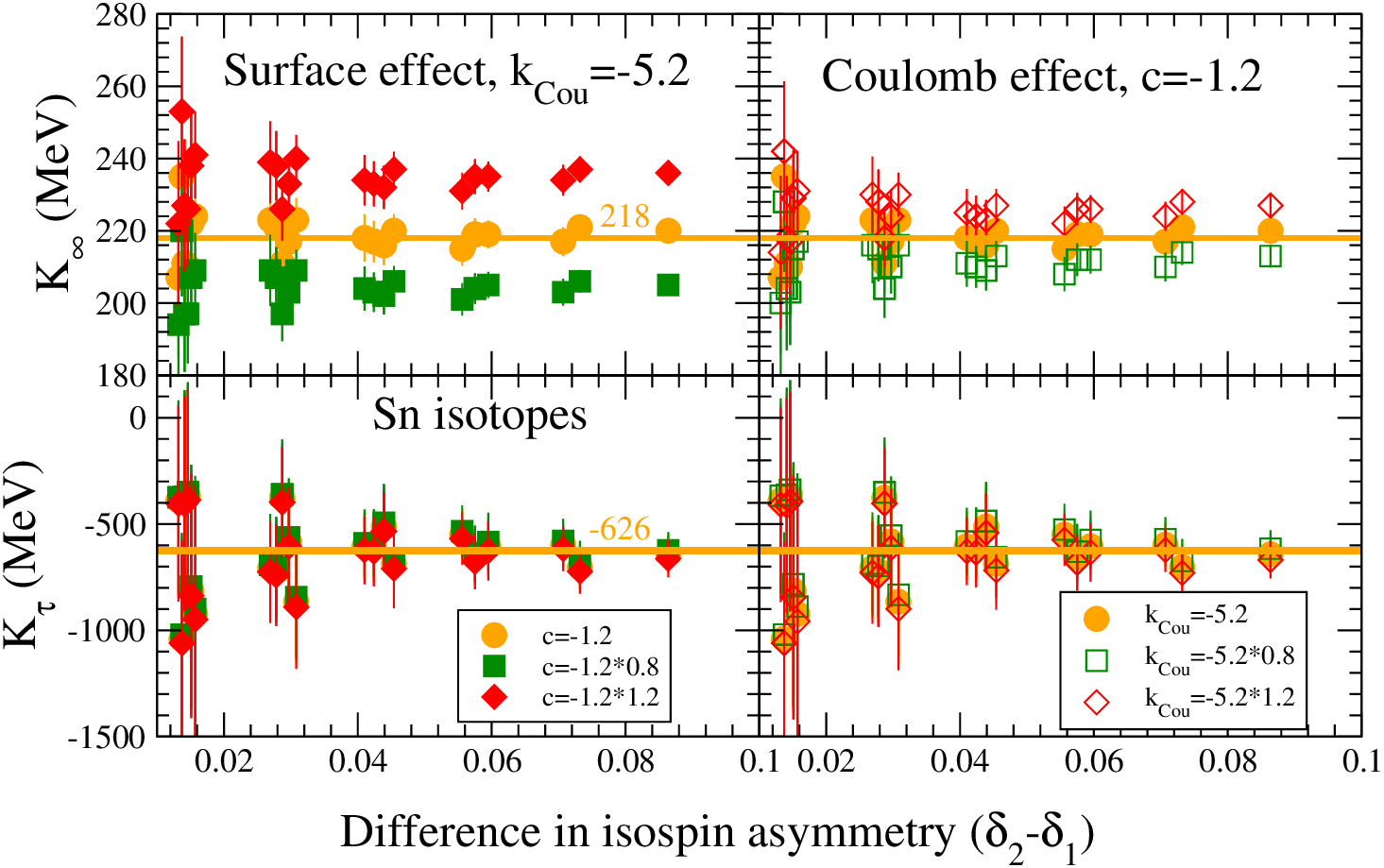}
 }
  \caption{(Color online) Variations of the \Kt (lower windows) and \Ki (upper windows) due to the variations of the surface parameter $c$ (left windows) and Coulomb parameter $K_{\rm Cou}$ (right windows), respectively, for the Sn isotopes. }\label{figure3}
\end{center}
\end{figure*}

It is interesting and necessary to check in more detail the consistency between the results of our differential analyses and those from the traditional integral analyses.
Shown in Fig. \ \ref{figure2} are the correlations between \Kt and \Ki with each point representing one nucleus-nucleus pair in the Cd or Sn isotope chain corresponding to the results shown in Fig. \ref{figure1}. Averaging over these results  is equivalent to performing a typical integral analysis, e.g., a multivariate $\chi^2$ fitting or Bayesian analysis. The solid lines are results of a $\chi^2$ fit to all points in the two isotope chains, separately.  The mean of the \Kt is $-625\pm 100$ MeV for the Cd isotopes and $-626\pm 188$ MeV for the Sn isotopes, respectively. The corresponding mean of \Ki is $213 \pm 3$ MeV for the Cd isotopes and $218 \pm 6$ MeV for the Sn isotopes, respectively. These mean values are in general agreement with the results of earlier $\chi^2$ analyses \cite{Garg18,Stone} of the same \Ka data using essentially identical surface and Coulomb parameters within error bars.

Interestingly, within the error bars of the mean values there is a clear anti-correlation between \Kt and $K_{\infty}$. It can be understood easily. With the surface and Coulomb parameters fixed, for a given \Ka value, the \Kt and \Ki is expected to be anti-correlated according to their relationship given in Eq. (\ref{KA-e}). Notice that the \Kt vs \Ki correlations for the Cd and Sn isotopes are almost in parallel in the direction of \Ki as they give approximately the same \Kt values but slightly different (about 5 MeV) \Ki values (notice the fine \Ki scale used). 

We emphasize that the error bars of the mean values of both \Kt and \Ki in the integral approaches, i.e., by averaging over all isotope pairs, are all much larger than those we found in the differential analyses of $^{106}$Cd-$^{116}$Cd and $^{112}$Sn-$^{124}$Sn pairs. Besides the advantage of largely cancelling the systematic errors in the differential analyses, another reason is that the $K_{\tau}\delta^2$ contribution to \Ka is very small even for the most neutron-rich nuclei available. For instance, with $\delta=0.2, K_{\tau}=-600$ MeV, $K_{\tau}\delta^2=-24$ MeV that is still only about 10\% of the acceptable $K_{\infty}$ values around 240 MeV. It is actually significantly less than the current uncertainty of about 40 MeV of the consensus value for $K_{\infty}$. A global $\chi^2$ fit to the \Ka data or Bayesian analysis of all \Ka data available thus can not reliably extract the value of $K_{\tau}$ from its small contribution relative to \Ki to the $K_{\rm A}$. In turn, the uncertainty of extracting the \Ki can not be better than $K_{\tau}\delta^2/K_{\infty}$ in the integral analyses of the \Ka data. On the contrary, the differential approach decouples completely the extractions of \Kt and \Ki for each isotope pair used. Only the \Kt and \Ki extracted independently for different isotope pairs along an isotope chain show an expected intrinsic correlation within their respective error bars.

\section{Effects of the surface and Coulomb parameters} \label{surf}
We have used above the known most probable values of $c=-1.2$ \cite{Pat02} and $K_{\rm Cou}=-5.2$  MeV \cite{Sag07}. It is also known that the Coulomb parameter is rather model independent \cite{Sag07,Cou2} while the calculations \cite{Blaizot,JPB2,Tre,JPB3,Myer1,Myer2} of the surface parameter $c$ show somewhat larger variations around $c\sim -1$. It is generally accepted that both the $c$ and $K_{\rm Cou}$ parameters have less than about $(10-20)\%$ uncertainties \cite{Garg18,Stone}. How do theses uncertainties affect the accuracies of extracting the \Kt and \Ki in the differential analyses? To answer this question, we have carried out systematic calculations by varying the two parameters independently by $\pm 20\%$ around their most probable values. 

As an example, shown in Fig. \ \ref{figure3} are the variations of the \Kt (lower windows) and \Ki (upper windows) due to the variation of the surface parameter $c$ (left windows) and Coulomb parameter $K_{\rm Cou}$ (right windows) for the Sn isotopes. Qualitatively, effects of varying the $K_{\rm Cou}$ and especially the parameter $c$ are much smaller on \Kt than on $K_{\infty}$. Quantitatively, for the $^{112}$Sn-$^{124}$Sn pair, changing the $c$ parameter by 40\% from $-1.2\times 0.8$ to $-1.2\times 1.2$ makes the \Kt change by about 6\% from $-624\pm 84$ MeV to $-663\pm 87$ MeV, while the \Ki changes by about 13\% from $205\pm 29$ MeV to $236\pm34$ MeV, respectively.
On the other hand, by changing the $K_{\rm Cou}$ by 40\% from $-5.2\times 0.8$ to $-5.2\times 1.2$, the \Kt change by about 8\% from $-616\pm 87$ MeV to $-669\pm 87$ MeV, while the \Ki changes by about
6\% from $213\pm 31$ MeV to $227\pm31$ MeV. Thus, the (6-8)\% uncertainty of \Kt due to the $\pm 20\%$ uncertainty of the surface parameter is much smaller than the approximately 14\% uncertainty due to the experimental errors of \Ka. While the (8-13)\% uncertainty of \Ki due to the $\pm 20\%$ uncertainty in the Coulomb parameter is compatible with that due to the experimental errors of the \Ka data. Thus, the remaining uncertainties of the surface and Coulomb parameters of about (10-20)\% have essentially no effect on the extraction of $K_{\tau}$.

The observed dependences of \Kt and \Ki on the variations of the surface and Coulomb parameters can be understood analytically by further examining the expressions of \Kt and \Ki in Eq. (2) and Eq. (3), respectively. Firstly, we examine effects of the parameter $c$. The c-dependent part of \Kt is
\begin{equation}
K_{\tau}\propto \frac{(1+cA_2^{-1/3})K_{\rm A_1}-(1+cA_1^{-1/3})K_{\rm A_2}}{(1+cA_2^{-1/3})\delta_1^2-(1+cA_1^{-1/3})\delta_2^2}.
\end{equation}
Because the parameter $c$ appears in all terms, its effect largely cancels out. In particular, for heavy nuclei $c/A^{1/3}\approx 0$, the \Kt then becomes independent of $c$, i.e.,
\begin{equation}
K_{\tau}\rightarrow \frac{K_{\rm A_1}-K_{\rm A_2}}{\delta_1^2-\delta_2^2}.
\end{equation}
While the c-dependent part of \Ki is
\begin{equation}
K_{\infty}\propto \frac{\delta_2^2K_{\rm A_1}-\delta_1^2K_{\rm A_2}}{c\cdot(\delta_2^2A_1^{-1/3}- \delta_1^2A_2^{-1/3})+\delta_2^2-\delta_1^2}.
\end{equation}
The parameter $c$ only appears in the first term of the denominator. Since both terms in the denominator are very small, a very small change in the parameter $c$ can thus lead to a large change in the value of $K_{\infty}$. This also implies that the surface properties of different nuclei may affect significantly the extraction of \Ki from the \Ka data as already noticed in the $\chi^2$ analyses in Ref. \cite{Stone}.

Similar analyses can be done to understand effects of the Coulomb parameter $K_{\rm Cou}$. More specifically,
\begin{equation}
K_{\tau}\propto -K_{\rm Cou}Z^2\frac{(1+cA_2^{-1/3})A_1^{-4/3}-(1+cA_1^{-1/3})A_2^{-4/3}}{(1+cA_2^{-1/3})\delta_1^2-(1+cA_1^{-1/3})\delta_2^2}.
\end{equation}
Again, the parameter $c$ has little effect as it appears in all terms. Considering $c A^{-1/3}\approx 0$ for heavy nuclei, the above expression reduces to
\begin{equation}
K_{\tau}\rightarrow -K_{\rm Cou}Z^2\times\frac{A_1^{-4/3}-A_2^{-4/3}}{\delta_1^2-\delta_2^2}.
\end{equation}
As heavy nuclei are more neutron rich in a given chain of isotopes, i.e., for $A_2>A_1$, $\delta_2>\delta_1$, the fraction in the above equation is always negative. Thus, one obtains $K_{\tau}\propto K_{\rm Cou}$. While the $K_{\rm Cou}$ itself is negative,
thus a larger negative $K_{\rm Cou}$ decreases the value of $K_{\tau}$ as seen in our numerical calculations. As the overall contribution of the Coulomb term to the $K_{\tau}$ is small, its variation causes little change in the final \Kt value. While for the Coulomb effect on $K_{\infty}$, a similar analysis leads to
\begin{equation}
K_{\infty}\propto -K_{\rm Cou}\frac{Z^2}{A_1^{4/3}A_2^{4/3}}\times\frac{\delta_2^2A_2^{4/3}-\delta_1^2A_1^{4/3}}{\delta_2^2-\delta_1^2}.
\end{equation}
Since the last fraction is always positive, thus $K_{\tau}\propto -K_{\rm Cou}$. Therefore, a larger negative $K_{\rm Cou}$ increases the value of $K_{\infty}$. Moreover, the above analysis clearly explains why the Coulomb parameter has opposite effects on extracting the \Kt and $K_{\infty}$ values.

\section{Summary and conclusions}
In summary, we emphasize the following aspects of our work
\begin{itemize}
\item
To our best knowledge, the differential approach we proposed here is original. The analytical expressions for \Ki and \Kt are derived from solving exactly two linear equations for two unknowns using the \Ka data as the only input. While the approach is very simple, its physics is sound. It also has no dependence on any nuclear many-body theories nor interactions. There is absolutely no fitting procedure involved, thus it does not suffer from the well-known problems in fitting the \Ka data using the Eq. (\ref{KA-e}).
\item
The \Ki and \Kt are directly extracted from the experimental \Ka data of any two nuclei in a given isotope chain. 
Besides showing that the new approach gives a \Ki consistent with its fiducial value  from the ``consensus" approach, and a \Kt that is much more accurate than what is available in the literature, we also ask for the first time the question which isotope pairs are most useful for extracting especially the \Kt at rare isotope beam facilities. Our answer to this question is expected to be useful for future experiments using rare isotopes to study the equation of state of neutron-rich matter.
\item
The so-called ``Soft Sn puzzle" (when one uses the interactions that correctly reproduce the GMR strength in $^{208}$Pb to calculate the GMR strengths in the Sn and Cd isotopes within the ``consensus" approach, the experimental values are always overestimated) has been alive for over 10 years. However, people have not really understood the underlying cause of the puzzle, leading to the conclusion that it is not feasible to simultaneously reproduce both the $^{208}$Pb and Sn's GMR data by the same interaction \cite{Jorge09} using the state-of-the-art theories within the ``consensus" approach. While we did not solve this puzzle in this work, we showed for the first time that the \Ki from analyzing the GMR data of Pb, Sn, Gd isotopes are not much different within the experimental error bars to indicate strongly a ``Soft Sn Puzzle". Quantitatively, the \Ki from Sn isotopes is only about 5 MeV smaller than that from the Pb isotopes. On the other hand, the \Kt from analyzing the $^{204,206,208}$Pb data is significantly higher (by $\sim 380$ MeV) than the converged asymptotic \Kt value at large isospin separations in analyzing the Sn and Gd isotopes, indicating strongly a ``Stiff Pb Phenomenon" \cite{Ehan}. To verify this further, differential analyses of future GMR data of more Pb isotopes will be very useful. The suggestion of having more Pb data was also made for addressing the same puzzle from a different perspective in Ref. \cite{Ehan}.

\item There are several caveats in our work. Firstly, if the leptodermous expansion of Eq. (\ref{KA-e}) is perfect, one expects the \Ki and \Kt extracted from all pairs of nuclei to be identical. In reality, this is of course not the case. As mentioned earlier, we expect the differential approach to work better for heavy nuclei along the same isotope chains. Indeed, the \Kt from most isotopes fall approximately on the same line at large isospin separations within still relatively large error bars, while the \Ki from different isotope chains especially the light nuclei scatter more broadly due to mostly the remaining (approximately $\pm 20\%$) uncertainty of the surface parameter $\mathrm{c}$. Secondly, to avoid introducing any model dependence in presenting their \Ka data, the experimentalists translated their original GMR observables to the ``experimental" \Ka data by using the experimentally measured charge radii instead of the matter radii which are inherently model dependent. This probably introduced a systematic error in the ``experimental" \Ka data and its effects have not been evaluated yet. We used the ``experimental" \Ka data as in all previous analyses in the literature. Thus, all results presented here should be understood within the context and with the cautions discussed above. Nevertheless, the importance and new physics revealed in our work can be clearly seen from comparing our approach and results with the traditional ones in the literature. We also emphasize that the focus of this work is a more accurate determination of $K_{\tau}$ for finite nuclei, while the \Ki for infinite nuclear matter just came out naturally consistent with its fiducial value that has not changed much since 1980.
\end{itemize}

In conclusion, we proposed a differential approach to analyze the incompressibilities of neutron-rich nuclei and investigated which nuclear pairs give the most accurate results using both the \Ka data and
analytically. The nucleus-nucleus pair having the largest difference in their isospin asymmetries in a given isotope chain is found to give the most accurate values of both \Kt and $K_{\infty}$ simultaneously. Applying this new approach to the \Ka data from RCNP, we found that  the $^{106}$Cd-$^{116}$Cd and $^{112}$Sn-$^{124}$Sn pairs give consistently the most accurate up-to-date \Kt values of $-616\pm 59$ MeV and $-623\pm 86$ MeV, respectively, largely independent of the remaining uncertainties of the surface and Coulomb parameters. These results can exclude many predictions based on various microscopic and/or phenomenological nuclear many-body theories in the literature.  We also studied the ``Stiff Pb Phenomenon" versus the ``Soft Sn Puzzle" and found that the former is favored. Thus, compared to the integral approach widely used in the literature, the differential analysis can reveal some interesting new physics underlying the incompressibilities of finite nuclei.

\section*{Acknowledgments.} We would like to thank Umesh Garg for many fruitful discussions.
BALI is supported in part by the U.S. Department of Energy, Office of Science, under Award Number DE-SC0013702 and the CUSTIPEN (China-U.S. Theory Institute for Physics with Exotic Nuclei) under the US Department of Energy Grant No. DE-SC0009971. WJXIE is supported in part by the Yuncheng University Research Project under Grant No. YQ-2017005 and The Scientific and Technological Innovation Programs of Higher Education Institutions in Shanxi under Grant No. 2020L0550.


\end{document}